\documentclass[aps,prc,twocolumn,showpacs,superscriptaddress,amsmath,amssymb,a4paper]{revtex4}
\usepackage{graphicx}
\usepackage{dcolumn}
\usepackage{bm}

\begin{document}

\title{Production of the $\omega$ meson in the $p d \rightarrow \mbox{$^3$He $\omega$}$ reaction at 1450 MeV and 1360 MeV}%
\author{K.~Sch\"onning}\affiliation{Department of Physics and Astronomy, Uppsala University, Uppsala, Sweden}
\author{Chr.~Bargholtz}\affiliation{Department of Physics, Stockholm University, Stockholm, Sweden}
\author{M.~Bashkanov}\affiliation{Physikalisches Institut der Universit\"at T\"ubingen, T\"ubingen, Germany}
\author{M.~Ber{\l}owski}\affiliation{So{\l}tan Institute of Nuclear~Studies, Warsaw and Lodz, Poland}
\author{D.~Bogoslawsky}\affiliation{Joint Institute for Nuclear Research, Dubna, Russia}
\author{H.~Cal\'en}\affiliation{Department of Physics and Astronomy, Uppsala University, Uppsala, Sweden}
\author{H.~Clement}\affiliation{Physikalisches Institut der Universit\"at T\"ubingen, T\"ubingen, Germany}
\author{L.~Demir\"ors}\affiliation{Institut f\"ur Experimentalphysik, Universit\"at Hamburg, Hamburg, Germany}
\author{C.~Ekstr\"om}\affiliation{The Svedberg Laboratory, Uppsala, Sweden}
\author{K.~Fransson}\affiliation{Department of Physics and Astronomy, Uppsala University, Uppsala, Sweden}
\author{L.~Ger\'en}\affiliation{Department of Physics, Stockholm University, Stockholm, Sweden}
\author{L.~Gustafsson}\affiliation{Department of Physics and Astronomy, Uppsala University, Uppsala, Sweden}
\author{B.~H\"oistad}\affiliation{Department of Physics and Astronomy, Uppsala University, Uppsala, Sweden}
\author{G.~Ivanov}\affiliation{Joint Institute for Nuclear Research, Dubna, Russia}
\author{M.~Jacewicz}\affiliation{Department of Physics and Astronomy, Uppsala University, Uppsala, Sweden}
\author{E.~Jiganov}\affiliation{Joint Institute for Nuclear Research, Dubna, Russia}
\author{T.~Johansson}\affiliation{Department of Physics and Astronomy, Uppsala University, Uppsala, Sweden}
\author{S.~Keleta}\affiliation{Department of Physics and Astronomy, Uppsala University, Uppsala, Sweden}
\author{O.~Khakimova}\affiliation{Physikalisches Institut der Universit\"at T\"ubingen, T\"ubingen, Germany}
\author{K.~P.~Khemchandani}%
\affiliation{Departamento de F\'isica Te\'orica and IFIC, Centro
Mixto Universidad de Valencia-CSIC, Institutos de Investigaci\'on de
Paterna, Aptd. 22085, 46071 Valencia, Spain.}%
\affiliation{Centro de F\'isica Computacional, Departamento de
F\'isica, Universidade de Coimbra, P-3004-516 Coimbra, Portugal}
\author{F.~Kren}\affiliation{Physikalisches Institut der Universit\"at T\"ubingen, T\"ubingen, Germany}
\author{S.~Kullander}\affiliation{Department of Physics and Astronomy, Uppsala University, Uppsala, Sweden}
\author{A.~Kup\'s\'c}\affiliation{Department of Physics and Astronomy, Uppsala University, Uppsala, Sweden}
\author{A.~Kuzmin}\affiliation{Budker Institute of Nuclear Physics, Novosibirsk, Russia}
\author{K.~Lindberg}\affiliation{Department of Physics, Stockholm University, Stockholm, Sweden}
\author{P.~Marciniewski}\affiliation{Department of Physics and Astronomy, Uppsala University, Uppsala, Sweden}
\author{B.~Morosov}\affiliation{Joint Institute for Nuclear Research, Dubna, Russia}
\author{W.~Oelert}\affiliation{Institut f\"ur Kernphysik, Forschungszentrum J\"ulich, Germany}
\author{C.~Pauly}\affiliation{Institut f\"ur Experimentalphysik, Universit\"at Hamburg, Hamburg, Germany}
\author{H.~Petr\'en}\affiliation{Department of Physics and Astronomy, Uppsala University, Uppsala, Sweden}
\author{Y.~Petukhov}\affiliation{Joint Institute for Nuclear Research, Dubna, Russia}
\author{A.~Povtorejko}\affiliation{Joint Institute for Nuclear Research, Dubna, Russia}
\author{R.~J.~M.~Y.~Ruber}\affiliation{Department of Physics and Astronomy, Uppsala University, Uppsala, Sweden}
\author{W.~Scobel}\affiliation{Institut f\"ur Experimentalphysik, Universit\"at Hamburg, Hamburg, Germany}
\author{R.~Shafigullin}\affiliation{Moscow Engineering Physics Institute, Moscow, Russia}
\author{B.~Shwartz}\affiliation{Budker Institute of Nuclear Physics, Novosibirsk, Russia}
\author{T.~Skorodko}\affiliation{Physikalisches Institut der Universit\"at T\"ubingen, T\"ubingen, Germany}
\author{V.~Sopov}\affiliation{Institute of Theoretical and Experimental Physics, Moscow, Russia}
\author{J.~Stepaniak}\affiliation{So{\l}tan Institute of Nuclear~Studies, Warsaw and Lodz, Poland}
\author{P.-E.~Tegn\'er}\affiliation{Department of Physics, Stockholm University, Stockholm, Sweden}
\author{P.~Th\"orngren Engblom}\affiliation{Department of Physics and Astronomy, Uppsala University, Uppsala, Sweden}
\author{V.~Tikhomirov}\affiliation{Joint Institute for Nuclear Research, Dubna, Russia}
\author{A.~Turowiecki}\affiliation{Institute of Experimental Physics, Warsaw, Poland}
\author{G.~J.~Wagner}\affiliation{Physikalisches Institut der Universit\"at T\"ubingen, T\"ubingen, Germany}
\author{C.~Wilkin}\affiliation{Physics and Astronomy Department, UCL, London, United Kingdom}
\author{M.~Wolke}\affiliation{Institut f\"ur Kernphysik, Forschungszentrum J\"ulich, Germany}
\author{J.~Zabierowski}\affiliation{So{\l}tan Institute of Nuclear~Studies, Warsaw and Lodz, Poland}
\author{I.~Zartova}\affiliation{Department of Physics, Stockholm University, Stockholm, Sweden}
\author{J.~Z{\l}oma\'nczuk}\affiliation{Department of Physics and Astronomy, Uppsala University, Uppsala, Sweden}
\collaboration{The CELSIUS/WASA Collaboration} \noaffiliation
\date{\today}

\begin{abstract}
The production of $\omega$ mesons in the $pd \to{}^3$He$\,\omega$
reaction has been studied at two energies near the kinematic
threshold, $T_p=1450$~MeV and $T_p=1360$~MeV. The differential cross
section was measured as a function of the $\omega$ cm angle at both
energies over the whole angular range. Whereas the results at
1360~MeV are consistent with isotropy, strong rises are observed near
both the forward and backward directions at 1450~MeV. Calculations
made using a two-step model with an intermediate pion fail to
reproduce the shapes of the measured angular distributions and also
underestimate the total cross sections.
\end{abstract}


\maketitle
%
%
\section{\label{sec:level1}Introduction}

Extensive studies of $\omega$ production in the $\pi^{-}p \to n\,
\omega$ reaction were carried out in the 1970s at the NIMROD
synchrotron from the kinematic threshold up to $\omega$ cm momenta of
$p_{\omega}^*=260$~MeV/$c$~\cite{Binnie,Keyne,Karami}. Though the
differential cross sections were found to be isotropic, the data also
showed a remarkable suppression of the $\omega$ production amplitude
near the threshold~\cite{Keyne,Karami}. In the work of Binnie
\emph{et al.}~\cite{Binnie}, where $\pi^{-}p\to n\,\eta'$,
$\pi^{-}p\to n\,\phi$ and $\pi^{-}p\to n\,\eta$ were studied with the
same apparatus, no similar effects were found in the $\eta'$ and
$\phi$ cases and a threshold enhancement, rather than a suppression,
was observed for the $\eta$. First, a final state interaction (FSI) effect,
where one of the pions from the $\omega$ decay scatters off the recoiling
neutron, was suggested as a possible explanation for the $\omega$
threshold suppression~\cite{Binnie}. This explanation was however tested and abandoned in Keyne \emph{et al.}~\cite{Keyne}, where an alternative
explanation were advanced in terms of a combination of $s$- and
$p$-wave resonances.

The data available for $\omega$ production in $pd \to{}^3$He$\,
\omega$ are rather scarce. The reaction was first studied near its
kinematic threshold at SATURNE~\cite{plouin}. The differential cross
section at $T_{p}=1450$~MeV ($p_{\omega}^* = 280$~MeV/$c$) measured
in the forward and backward regions with the SPESIII
spectrometer~\cite{kirchner} showed clear anisotropy in
$\theta_{\omega}^*$, with strong peaking at extreme angles. This is
in contrast to the angular distributions observed in $\pi^{-}p\to n\,
\omega$, which remain flat up to at least $p_{\omega}^* \approx
200$~MeV/$c$~\cite{Karami}.

Wurzinger \emph{et al.}\ measured the $pd\to{}^3$He$\,\omega$ cross
section at $\theta_{\omega}^*=180^\circ$ as a function of energy
using the SPESIV spectrometer at SATURNE~\cite{wurzinger} and found a
suppression in the production amplitude near threshold similar to the
one observed in the $\pi^{-}p \to n\, \omega$
reaction~\cite{Binnie,Keyne,Karami}. These authors also described
their data in terms of the FSI hypothesis. However, the correctness
of this interpretation has been questioned for both reactions in
Ref.~\cite{hanhart1,hanhart2}, where effects associated with the
$\omega$ width have been stressed.

The production of heavy mesons in $pd$ collisions has been studied
theoretically in a two-step
model~\cite{gorancolin1,gorancolin2,kondrauzi,we3,we3_sec,we4,stenmark2003,ulla},
which first involves the production of a light meson in the
interaction between the incident and one of the target nucleons. In
the second step, the light meson interacts with the other target
nucleon to create the observed heavy meson. This procedure allows the
large momentum transfer to be shared between the nucleons. The
predictions of this model have been evaluated for the specific
$pd\to{}^3$He$\,\omega$ reaction~\cite{kondrauzi}, but only at the
backward angles where data existed~\cite{wurzinger}, and no previous
attempts have been made to calculate the full angular distribution.

In this paper, the results of the measurements of $\omega$ production
in the $pd\to{}^3$He$\,\omega$ reaction by the CELSIUS/WASA
collaboration are presented at two different beam kinetic energies,
$T_{p}=1450$~MeV and $T_{p}=1360$~MeV. These correspond,
respectively, to $\omega$ cm momenta of $p^{*}_{\omega}=280$~MeV/$c$
and $p^{*}_{\omega}= 144$~MeV/$c$, \emph{i.e.}, excess energies of
63~MeV and 17~MeV. We have previously reported data on the angular
distribution of the $\omega$ decay plane that show that the $\omega$
is produced largely unpolarized~\cite{Karin_PLB}, which is in stark
contrast to the almost complete polarization of the $\phi$ from the
analogous $pd\to{}^3$He$\,\phi$ reaction~\cite{MOMO}. We now provide
the corresponding results on the differential cross sections that
have been measured over the full angular range. These new data
constitute an important input to the intriguing question of how heavy
mesons are produced in few-body collisions and, in particular,
whether the reaction can be usefully viewed as a sequential two-step
process.

The outline of this paper is as follows. We start by introducing the
WASA detector before going through the analysis in some detail. This
involves principally the $^3$He identification, event selection,
background, and normalization procedure. After describing the
acceptance corrections, we show our results and compare them to
calculations that have been performed in terms of a two-step model.
%
%
\section{The CELSIUS/WASA experiment}

\begin{figure}
\begin{center}
\includegraphics[width=0.95\columnwidth]{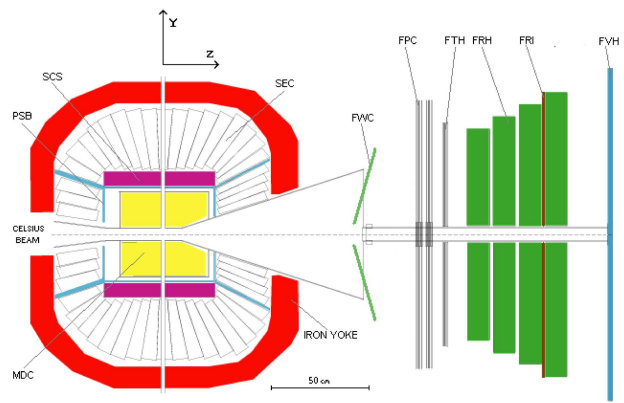}
\caption{Side view of the CELSIUS/WASA detector setup. The Central
Detector built around the interaction point (to the left) is
surrounded by an iron yoke. The layers of the Forward Detector are
shown in the right-hand side. The CELSIUS beam pipe runs horizontally
and the target pellets are injected through the vertical pipe. The
abbreviations in the figure are explained in the text.}
\label{fig:wasa4pi}
\end{center}
\end{figure}

The measurements of the $pd\to{}^3$He$\, \omega$ reaction were
carried out at the The Svedberg Laboratory in Uppsala, Sweden, using
the WASA detector~\cite{Zabierowski} which, until June 2005, was an
integral part of the CELSIUS storage ring. A deuterium pellet
target~\cite{Ekstrom, Nordhage} was used in combination with a proton
beam. The $^3$He were detected in the WASA forward detector (FD),
which covers laboratory polar angles from $3^\circ$ to $18^\circ$
with respect to the beam direction. The forward detector consists of
a sector-like window counter (FWC) for triggering, a proportional
chamber (FPC) for precise angular information, a trigger hodoscope
(FTH) for triggering and off-line particle identification, a range
hodoscope (FRH) for energy measurements, particle identification and
triggering, an intermediate hodoscope (FRI) for improved track 
reconstruction and neutron detection and finally a veto hodoscope (FVH) 
for triggering. All FD
elements, except the FPC, are made of plastic scintillators.

Mesons and their decay products are detected mainly in the central
detector (CD), which consists of the Plastic Scintillating Barrel
(PSB), the Mini Drift Chamber (MDC), and the central Scintillating
Electromagnetic Calorimeter (SEC) that is made of CsI crystals.
Charged particles, such as pions from $\omega$ decay, are
distinguished from neutral particles by their signals in the PSB,
which also provides angular determination from $24^\circ$ to
$159^\circ$. The momenta of charged particles are estimated by
tracking in a magnetic field from the SuperConducting Solenoid (SCS)
using information from the MDC. The SEC measures angles and energies
of photons from meson decays in the polar angle range from $20^\circ$
to $169^\circ$. A schematic overview of the CELSIUS/WASA detector
setup is shown in Fig.~\ref{fig:wasa4pi}.

A special trigger was developed to select $^3$He events, which are
characterized by one high-energy-deposit hit in the FWC and one hit
that overlaps in the azimuthal angle in one of the consecutive
detectors.
%
%
\section{Analysis}
\subsection{$^3$He identification}

\begin{figure}
\begin{center}
\includegraphics[width=0.95\columnwidth]{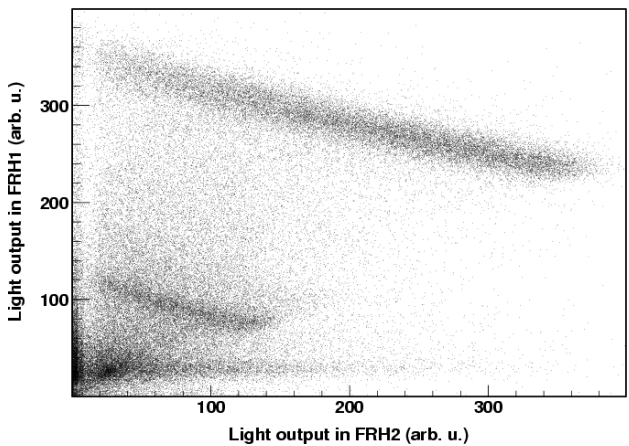}
\caption{The scintillation light output in the first layer of the FRH
\textit{versus} the light output in the second layer for particles
stopping in the second layer. The uppermost band corresponds to
$^3$He ions, below which is a band of stopping protons. The intense
spot at the bottom left corresponds to minimum ionizing particles.}
\label{fig:frh1frh2}
\end{center}
\end{figure}

The $^3$He ions are identified in the FD by the $\Delta E-E$ method.
For this the scintillation light output in the detector layer where
the particle stops is compared to that in the preceding layer. The
particles then show up in different bands, as illustrated in
Fig.~\ref{fig:frh1frh2}, where the uppermost one represents the
$^3$He. For these candidates the scintillation light output is
converted to energy deposit, taking into account the light
quenching~\cite{birks}. This allows the corresponding range to be
estimated, which is added to the amount of material traversed by the
particle in the preceding detector layers to give the total range.
Energy-range tables then allow the initial kinetic energy to be
estimated. The expected light output in all the detector layers
traversed is calculated and compared to the measured one by
evaluating the $\chi^2$, according to
\begin{equation}
\chi^2=\sum_{i=1}^N(\Delta L_i^m-\Delta L_i^c)^2/\sigma_i^2\,,
\label{eq:chisquare}
\end{equation}
where $i$ is the layer number, $N$ the number of layers traversed,
$\Delta L_i^m$ the measured light output, $\Delta L_i^c$ the
calculated light output, and $\sigma_i$ the light output uncertainty
in a given layer. The particle is then considered to be properly
identified if the calculated $\chi^2$ does not exceed
$\chi^2_{\rm{max}}=6.0$. For more details, see Ref.~\cite{jozefPID}.

%
%
\subsection{Event selection}
In this work, we focus primarily on the $\omega \to \pi^0\pi^+\pi^-$
decay channel because the large branching ratio (BR=89.1$\%$) gives
the high statistics required for the extraction of angular
distributions. However, a parallel analysis of the $\omega \to
\pi^0\gamma$ decay channel (BR=8.7$\%$) is a valuable tool for
checking that cut efficiencies and other experimental biases are
under control.
%
%
\subsubsection{$pd\to{}^3\!$He$\,\omega$,\ $\omega \to \pi^0\pi^+\pi^-$ events}
\label{sec:pipimpi0}%
To select $\omega \to \pi^0\pi^+\pi^-$ events, we require the $^3$He
ion to be measured in the FD with well defined energy and angle,
which means that the $^3$He has to stop within the FRH. The
geometrical acceptance of the FD is 95\% at 1450~MeV and 78\% at
1360~MeV. The main event loss is due to the $^3$He emitted at small
angles that escape down the beam pipe. The detection efficiency of
the FD is further reduced by nuclear interactions in the detector
material. The efficiency of detecting a $^3$He from
$pd\to{}^3$He$\,\omega$ in the FD, as deduced from Monte Carlo
simulations including geometry and detector material responses, is 61\% and
54\% at 1450 and 1360~MeV, respectively.

In addition to the $^3$He selection, at least two photons are
required in the SEC. Furthermore, one photon pair must have an
invariant mass that does not differ from that of the $\pi^0$ by more
than 45~MeV/$c^2$, and the missing mass of the $^3$He$\,\pi^0$ system
must be larger than 250~MeV/$c^2$, \emph{i.e.}, twice the pion mass,
after taking the resolution into account. Finally, two or more hits
are needed in the PS and the total energy deposit in the SEC must not
exceed 900 MeV. This rejects candidates with photon signals from
accidental events.

\begin{figure}
\begin{center}
\includegraphics[width=0.95\columnwidth]{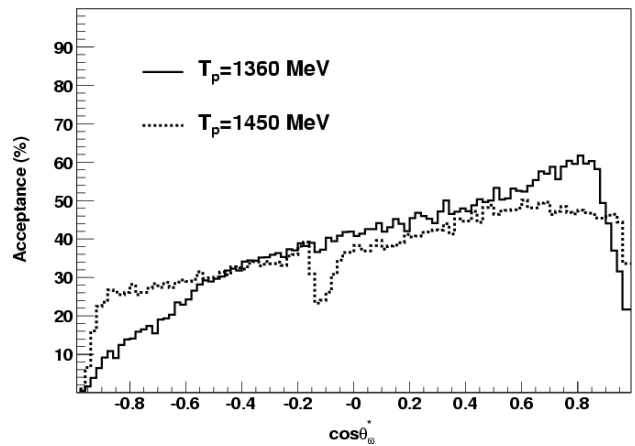}
\caption{The acceptance for the $pd\to{}^3$He$\,\omega,\ \omega\to
\pi^+\pi^-\pi^0$ reaction as a function of $\cos\theta_{\omega}^*$ at
1360~MeV (solid line) and 1450~MeV (broken line) after applying the
cuts explained in the text. The dip in the 1450~MeV distribution
corresponds to the $^3$He that stop between the first and second
layer of the FRH.} \label{fig:acc}
\end{center}
\end{figure}

These constraints give an overall acceptance of 35\% at 1450~MeV and
34\% at 1360~MeV. The differential acceptance is shown in
Fig.~\ref{fig:acc} for the two energies as a function of
$\cos\theta_{\omega}^*$.

%
%
\subsubsection{$pd\to{}^3\!$He$\,\omega,\ \omega \to \pi^0\gamma$ events}
\label{sec:pi0g}%
All final state particles, \emph{i.e.}, the $^3$He and three photons,
can be measured with high acceptance in the $\omega \to \pi^0\gamma$
case. Each event is fully reconstructed, which ensures that the
kinematical constraints are fulfilled. Though this leads to a cleaner
sample than the three-pion channel, the low branching ratio
(BR=8.7$\%$) yields statistics that are insufficient to provide
angular distributions.

Having identified a $^3$He and three photons, we demand that one
photon pair has an invariant mass close to that of the $\pi^0$,
$|IM(\gamma\gamma)-m_{\pi^0}| < 45$~MeV/$c^2$, and that the invariant
mass of all three photons is larger than 600~MeV/$c^2$. The
missing-mass-squared of the $^3$He$\gamma\gamma\gamma$ system must
not exceed $(100)^2\,$MeV$^2/c^4$. We also require that the
difference between the directions of the missing momentum of the
$^3$He and that of the 3$\gamma$ system be smaller than 20$^\circ$.
As in the three-pion case, a limit is placed on the total energy
deposit in the SEC, which may not exceed 1200 MeV. Finally, a
coplanarity cut is applied: $160^\circ<|\phi_{\text{lab}}(^3$He$)-
\phi_{\text{lab}}(3\gamma)|<200^\circ$. These constraints reject very
effectively the contribution from accidentals and events originating
far from the target region. They give acceptances of 19\% and 18\% at
1450~MeV and 1360~MeV, respectively.
%
%
\subsection{Backgrounds}
\label{sec:background}%

In the three-pion case, the principal background is direct pion
production, \emph{i.e.}, $pd\to{}^3$He$\,\pi^0\pi^+\pi^-$, which has
exactly the same signature as $\omega$ production. The acceptance for
this background, assuming phase space production, is 31\% at 1450~MeV
and 35\% at 1360~MeV.

Both $\omega$ and direct $3\pi$ events are produced mainly through
the interaction of the beam with the pellet target. However, such
reactions can also occur through interactions of the beam with the
rest gas, the beam halo with the rest gas, and the beam halo with the
beam pipe. Rest gas is produced when the pellets are vaporized by the
beam~\cite{nordhage2} or when the pellets are not properly captured
in the pellet beam dump.  In the present work, we found that around
30\% of the data was not produced from the region of the pellet
target. This is in line with the figures obtained from a recent
measurement of the $pd\to{}^3$He$\,\eta$ reaction at $T_{p} =
893$~MeV~\cite{stockholm}. The vaporization of the pellet is expected
to be higher at lower energies~\cite{nordhage2}, but the amount of
rest gas also depends upon other time-dependent experimental
conditions, such as the beam-target overlap, the efficiency of the
vacuum pumps, and the performance of the pellet dump. Since the
energy reconstruction procedure assumes that all particles come from
a well defined interaction point, the measured energy and momentum of
particles produced far from the pellet target will have larger
uncertainties and thus degrade the missing-mass resolution. However,
the width of the $\omega$ peak agrees well with Monte Carlo
simulations for a well defined interaction point, which means that
most events come either from the target or from the close vicinity.
The latter can be treated as good events that do not significantly
affect the quality of the data sample.

Monte Carlo simulations show that other channels, such as
$pd\to{}^3$He$\,\pi^+\pi^-$, $pd\to{}^3$He$\,\pi^0\pi^0$ and
$pd\to{}^3$He$\,\pi^0\pi^0\pi^+\pi^-$, give negligible contributions
to the background and will be ignored.

The main background channel in the $\pi^0\gamma$ sample comes from
the $pd\to{}^3$He$\,\pi^0\pi^0$ reaction, where one of the photons
escapes detection. Assuming phase space production, about 1.8\% of
the $2\pi^0$ events survive the cuts optimized for $\pi^0\gamma$
selection at 1450~MeV and 1.4\% at 1360~MeV. Although these are tiny
fractions, the two-pion production cross section is large and the
$\omega \to \pi^0\gamma$ branching ratio small. Hence it is expected
that the numbers of events surviving the cuts should be of the same
order of magnitude for the signal and background.

Interactions between the beam halo and the beam pipe, as well as
chance coincidences, may also contribute to the background. However,
we show in Sec.~\ref{sec:results} that they only do so to a very
limited extent.
%
%
\subsection{Normalization}
\label{sec:normalisation}%

The luminosity depends on the beam intensity and the pellet rate,
which are monitored and recorded during the data taking. However, the
luminosity also depends on the overlap between the beam and target,
which cannot be measured directly. It is also not constant during a
run or even over a cycle of data taking. Furthermore, as previously
mentioned, as many as 30\% of the events may result from interactions
away from the target. This means that the luminosity must be
evaluated by normalizing the number of events of some subsidiary
reaction to a well known cross section. Quasi-elastic $pp$ scattering
might be a good choice because the cross section is well known and
large statistics are available. However, these data were collected
with a different trigger from the $^3$He events and the relative
trigger efficiency is not known with sufficient precision. It is
therefore preferable to use a reaction with a $^3$He in the final
state, where the trigger was the same as for $pd\to{}^3$He$\,\omega$.

The differential cross section for the $pd\to{}^3$He$\,\eta$ reaction
was measured at a few very backward angles with the SPESIV
spectrometer at 1450, 1350, and 1250~MeV~\cite{berthet}. It was also
measured over a larger angular range at 1450~MeV with the SPESIII
spectrometer~\cite{kirchner}. The two data sets agree well in the
region of overlap. WASA has collected clean $pd\to{}^3$He$\,\eta,\
\eta \to \gamma\gamma$ samples at 1450 and 1360~MeV. In addition, we
analyzed the $pd\to{}^3$He$\,\eta$ reaction \emph{via} the $\eta \to
\pi^0\pi^+\pi^-$ decay channel, which provided a cross check of the
cut efficiency. There is a good agreement between the number of
$\eta$ mesons in the $2\gamma$ and $3\pi$ samples~\cite{karinthesis}.
This shows that the effects of the cuts are well understood and that
the yield of the events coming from outside the target is about the
same in the $pd\to{}^3$He$\,\eta,\ \eta \to \gamma\gamma$ case as
when the cuts are optimized for selecting a $^3$He\,$\pi^0\pi^+\pi^-$
final state (including $\pi^0\pi^+\pi^-$ production \emph{via} $\eta$
and $\omega$).

The integrated luminosity $L$ is calculated from:
\begin{equation}
\frac{\Delta N}{\Delta \Omega}=\frac{d\sigma}{d\Omega}L
\end{equation}
where $\Delta N$ is the acceptance-corrected number of $\eta$
candidates in the angular region
$\Delta\Omega=2\pi\Delta(\cos\theta_{\eta}^*)$, and $d\sigma/d\Omega$
is the differential cross section at this angle.

\begin{figure}
\begin{center}
\includegraphics[width=0.95\columnwidth]{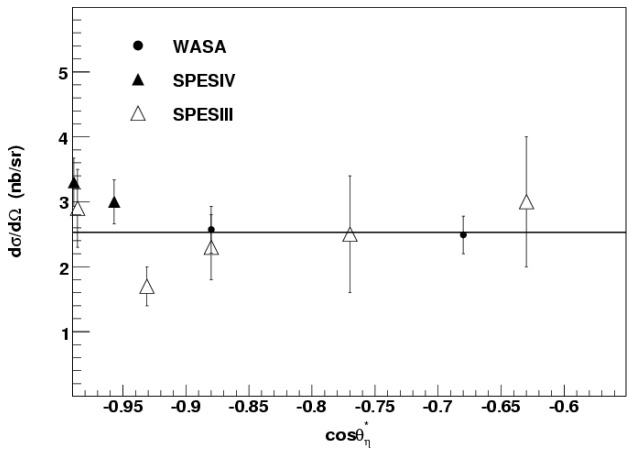}
\caption{Differential cross section data for the
$pd\to{}^3$He$\,\eta$ process at 1450~MeV, which has been used for
the evaluation of the luminosity. The filled and unfilled triangles
show data from SPESIV~\cite{berthet} and SPESIII~\cite{kirchner},
respectively. The line is a constant fitted to both sets of data
points. The filled circles represent the WASA data after
normalization.} \label{fig:norm1450}
\end{center}
\end{figure}

At 1450~MeV, there is a region $-0.99 < \cos\theta_{\eta}^* < -0.6$
where two WASA points overlap with a set of SPESIII and SPESIV points
(see Fig.~\ref{fig:norm1450}). In this region, the angular
distribution is approximately isotropic. A fit to the combined
SPESIII and SPESIV points to a constant gives a
$\chi^2/\rm{ndf}=1.8$. The resulting cross section, $d\sigma/d\Omega
= 2.53\pm0.19$~nb/sr, where the statistical, systematic and
normalization uncertainties of each point have been taken into
account. A fit to the two WASA acceptance-corrected numbers of counts
to a constant has an uncertainty of 9\% and this contributes to the
12\% overall uncertainty in the resultant luminosity.

\begin{figure}
\begin{center}
\includegraphics[width=0.95\columnwidth]{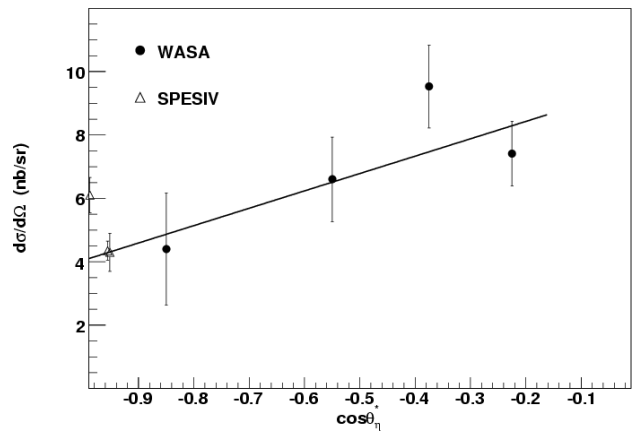}
\caption{Differential cross section data for the
$pd\to{}^3$He$\,\eta$ reaction used for the evaluation of the
luminosity at 1360~MeV. The unfilled triangles are data from
SPESIV~\cite{berthet} taken at $T_{p}=1350$~MeV; the point at
$\cos\theta_{\eta}^*=-0.96$ is used for normalization. The filled
circles are WASA data taken at 1360~MeV and the line is the result of
a linear fit to these.} \label{fig:norm1360}
\end{center}
\end{figure}

The poorer $pd\to{}^3$He$\,\eta$ database at 1360~MeV makes the
situation more difficult at this energy and we have to rely on the
few SPESIV points taken at $T_p = 1350$~MeV in the near-backward
region~\cite{berthet}. At 1250~MeV and lower, this data set extends
slightly further in angle and from this it is clear that, in addition
to the very sharp backward peak, the angular distributions show a
minimum for $\cos\theta_{\eta}^*\approx -0.96$ before rising less
steeply at more forward angles. Though the four WASA points in the
backward hemisphere shown in Fig.~\ref{fig:norm1360} are consistent
with such a small rise, they do not overlap with the SPESIV data so
that a linear fit has been made to extrapolate to the average of the
two points in the $\cos\theta_{\eta}^*=-0.96$ region. It could be
checked from the SPESIV data at 1450~MeV that the energy dependence
in this region is negligible compared to a total normalization
uncertainty of 29\%, of which 27\% arises from that in the linear
extrapolation.

%
%
\section{Results}
\label{sec:results}
\subsection{$T_p=1450$~MeV}
The $\omega$ mesons are identified from the peak in the $^3$He
missing-mass distribution. The number of $\omega$ events is extracted
by subtracting the background in two ways. The first is by fitting
the background to a phase space Monte Carlo simulation of
$pd\to{}^3$He$\,\pi^0\pi^+\pi^-$ data. The second is by fitting the
data as a Gaussian peak sitting on a polynomial background. The
number of background events can then be estimated by integrating the
polynomial. The central value and the width of the Gaussian peak
obtained in the fit provide a quality control.

\begin{figure}[hbt]
\begin{center}
\includegraphics[width=0.95\columnwidth]{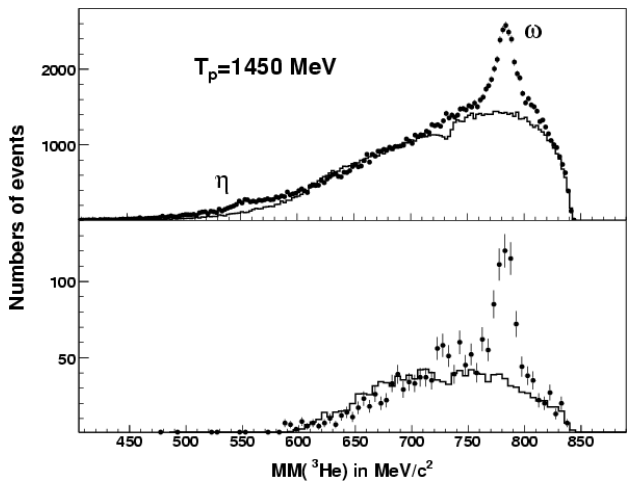}
\caption{Upper panel: Points are the distribution of missing mass of
the $^3$He, MM($^3$He), for all events at 1450~MeV fulfilling the
constraints optimized for $pd\to{}^3$He$\,\omega,\ \omega \to
\pi^0\pi^+\pi^-$ selection. The histogram shows the corresponding
phase space simulations of the $pd\to{}^3$He$\,\pi^0\pi^+\pi^-$
channel. Lower panel: MM($^3$He) distribution for events fulfilling
cuts optimized for selection of the $pd\to{}^3$He$\,\omega,\ \omega
\to \pi^0\gamma$ channel. The histogram is the phase space simulation
of $pd\to{}^3$He$\,\pi^0\pi^0$.} \label{fig:mm1450}
\end{center}
\end{figure}

The difference in the number of $\omega$ events extracted by
subtracting the background in these two ways gives the largest
contribution to the systematic uncertainty. Applying the constraints defined in
Sec.~\ref{sec:pipimpi0} gives a sample with a $^3$He missing mass
distribution shown in Fig.~\ref{fig:mm1450}. After subtracting the
background, we are left with 9900 $\pm$ 700 candidates for the
$\omega \to \pi^0\pi^+\pi^-$ channel.

To verify that the systematics are under control, we made a cross
check using the $\omega \to \pi^0\gamma$ channel. On the basis of the
evaluated acceptances and the known branching ratios, both summarized
in Table~\ref{tab:results}, we expect $520\pm40$ events from $\omega
\to \pi^0\gamma$ in the sample. Selecting the events according to
method described in Sec.~\ref{sec:pi0g} we get a number of
$420\pm50$, which is fairly consistent.

\begin{table}[hbt]
\caption{Summary of the numbers of $\omega$ candidates found at
different energies for the two decay channels.\vspace{2mm}
\label{tab:results}}
\begin{tabular}{ccccc}
\hline
Channel & $T_{p}$ & Acceptance & BR (\%) & Number of \\
        & (MeV)   & (\%)       & (\%)    & candidates\\
\hline
$\omega \to \pi^+\pi^-\pi^0$ & 1450 & 35 & 89.1 & 9900$\pm$700 \\
\hline
$\omega \to \pi^0\gamma$ & 1450 & 19 & 8.7 & 420$\pm$50 \\
\hline
$\omega \to \pi^+\pi^-\pi^0$ & 1360 & 34 & 89.1 & 1800$\pm$200 \\
\hline
$\omega \to \pi^0\gamma$ & 1360 & 18 & 8.7 & 80$\pm$20 \\
\hline
\end{tabular}
\end{table}

The good match between the background continuum and the simulated
$pd\to{}^3$He$\,\pi^0\pi^+\pi^-$ in Fig.~\ref{fig:mm1450}, the good
agreement between the $\omega$ width in Monte Carlo and data, and the
consistency between the results for the $3\pi$ and $\pi^0\gamma$
channels gives us confidence that the contribution from background
not coming from reproducible physical reactions may be safely
neglected.

\begin{figure}[hbt]
\begin{center}
\includegraphics[width=0.95\columnwidth]{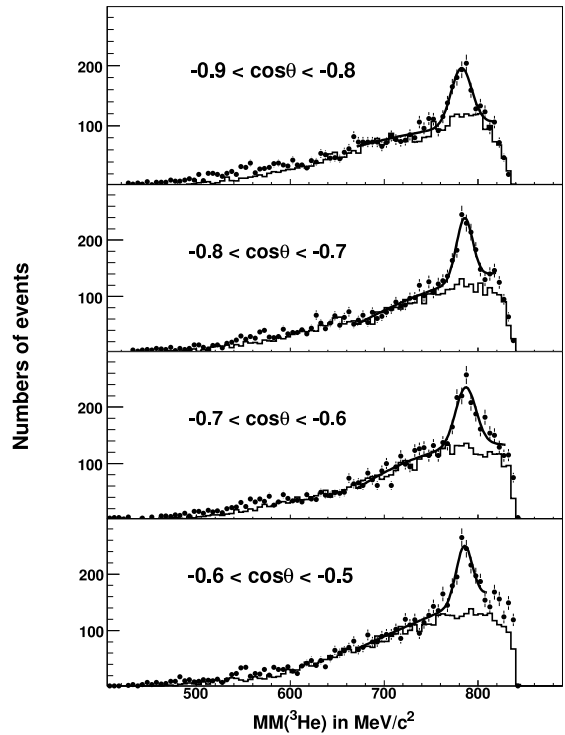}
\caption{Data taken in four angular regions at 1450~MeV, with cuts
optimized for the $pd\to{}^3$He$\,\omega,\ \omega \to
\pi^0\pi^+\pi^-$ selection. Phase space simulations of
$pd\to{}^3$He$\,\pi^0\pi^+\pi^-$ production are also shown, as well
as a fitted Gaussian peak on a polynomial background.}
\label{fig:mm_ang1450}
\end{center}
\end{figure}

In order to extract the differential cross section for the $3\pi$
sample, we divided the data into $\cos\theta_{\omega}^*$ bins and
plotted separate $^3$He missing-mass distributions, some examples of
which are shown in Fig.~\ref{fig:mm_ang1450}. The number of $\omega$
events was then obtained by subtracting the background, which was
estimated in the two ways described for the total sample. The
systematic uncertainty from background subtraction varies between 5\%
and 20\%. Adding the $\omega$ candidates in each bin gives 9090,
which is in good agreement with the value quoted in
Table~\ref{tab:results}. The numbers were corrected for acceptances
that were derived from the phase space Monte Carlo simulations. Since
the angular distribution is highly anisotropic, a fitted second
degree polynomial was used as input. This reduced the total
acceptance from 35\% to 33\%, though the effect for individual bins
was negligible. This shows that the model dependence of the acceptance on
angle is very weak. Finally we converted the numbers
of $\omega$ mesons to cross sections by using the integrated
luminosity discussed in Sec.~\ref{sec:normalisation}.

\begin{figure}
\begin{center}
\includegraphics[width=0.95\columnwidth]{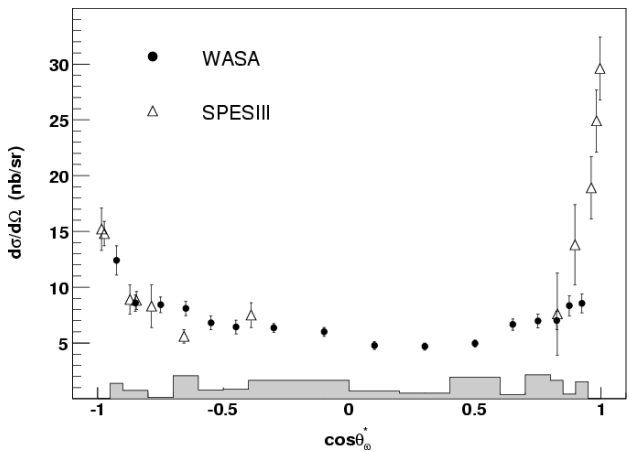}
\caption{$pd\to^3$He$\,\omega$ differential cross section at
1450~MeV. The error bars on the WASA data (filled circles) show the
statistical uncertainties while the grey histogram shows the
systematical uncertainties due to background subtraction and
acceptance correction. In addition there is an overall normalization
uncertainty of 12\%. The SPESIII results~\cite{kirchner} are shown by
the open triangles.} \label{fig:ome_ang_exp1450}
\end{center}
\end{figure}

The resulting $pd\to{}^3$He$\,\omega$ angular distribution is shown
in Fig.~\ref{fig:ome_ang_exp1450} along with the values obtained at
SPESIII~\cite{kirchner}. The two data sets agree in the backward
direction, which lends confidence to our normalization procedure.
However, although the WASA data do show a slight increase towards the
forward direction, the sharp peak observed in Ref.~\cite{kirchner} is
not confirmed.

Fitting the angular distribution with Legendre polynomials results in
an integrated cross section of
$\sigma_{\text{tot}}=83.6\pm1.5\pm2.2$~nb, where the first
uncertainty is statistical and the second systematic. In addition,
however, there is an uncertainty of 12\% that comes from the
luminosity determination.
%
%
\subsection{$T_p=1360$~MeV}

\begin{figure}[hbt]
\begin{center}
\includegraphics[width=0.95\columnwidth]{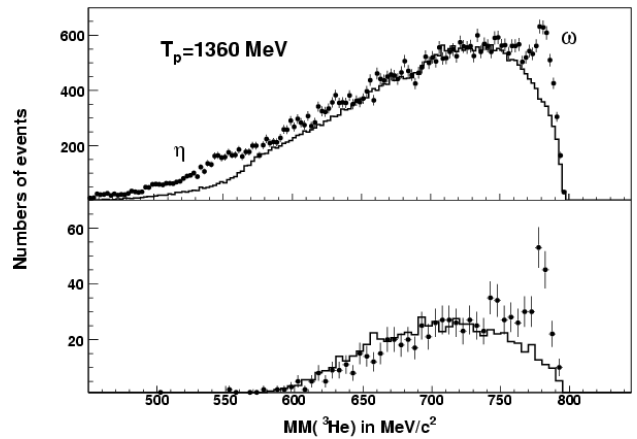}
\caption{MM($^3$He) data at 1360~MeV for both the $3\pi$ and
$\pi^0\gamma$ selection with experiment and simulation, as described
for the 1450~MeV data of Fig.~\ref{fig:mm1450}.} \label{fig:mm1360}
\end{center}
\end{figure}

The analysis is more difficult at 1360~MeV than at the higher energy.
The finite $\omega$ width ($\Gamma = 8.44$~MeV/$c^2$) leads to an
asymmetric peak close to threshold since the high mass tail of the
$\omega$ meson is then partially suppressed. Secondly, the background
subtraction is more difficult since the multipion continuum ends
under the $\omega$ peak (see Fig.~\ref{fig:mm1360}). Thirdly, the
signal-to-background ratio is small. To add to these difficulties,
fewer data were taken at 1360~MeV than at 1450~MeV due to beam-time
constraints. Higher systematic and statistical uncertainties are
therefore to be expected at this energy.

Selecting data according to the methodology of
Sec.~\ref{sec:pipimpi0} gives the data sample shown in the upper
panel of Fig.~\ref{fig:mm1360} which, after background subtraction,
leaves $1800\pm200$ $\omega$ events. Using the information given in
Table~\ref{tab:results}, we then estimate the number of $\omega \to
\pi^0\gamma$ events should be $90\pm10$. This is to be compared to
$80\pm20$ events obtained by subtracting the background in the lower
panel of Fig.~\ref{fig:mm1360}. The numbers from the two decay
channels are thus quite consistent.

\begin{figure}[hbt]
\begin{center}
\includegraphics[width=0.95\columnwidth]{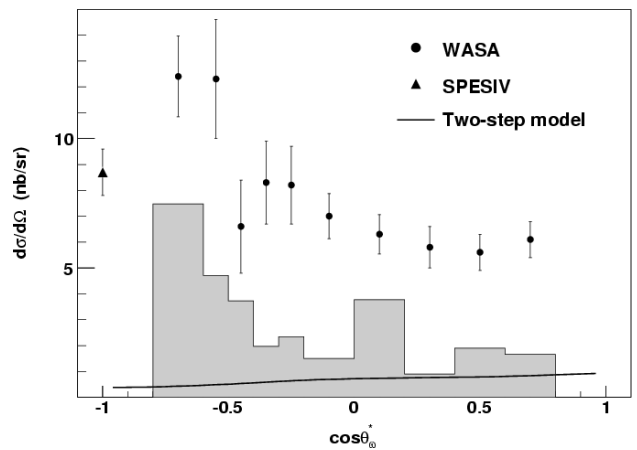}
\caption{Angular distributions for the $p d \to ^3$He$\,\omega$
reaction at 1360~MeV, as described in Fig.~\ref{fig:ome_ang_exp1450}
for the 1450~MeV results. In addition to the systematic uncertainties
shown by the grey histogram, there is an uncertainty from normalisation 
of 29\%. The triangle shows the differential
cross section obtained by interpolating the SPESIV
data~\cite{wurzinger}. The solid line represents the results of the
model calculations described in Sec.~\ref{sec:modelcalc}.}
\label{fig:ang1360}
\end{center}
\end{figure}

The extraction of the angular distribution was carried out in the
same way as at 1450~MeV. The sum of the events in each angular bin
(1600) agrees well with the 1800 found from Fig.~\ref{fig:mm1360}.
The angular distribution shown in Fig.~\ref{fig:ang1360} is
consistent with isotropy and, within the total uncertainties, agrees
with the point at 1360~MeV obtained by interpolating SPESIV
results~\cite{wurzinger}.\vspace{1mm}

It was suggested in Ref.~\cite{wurzinger} that rescattering of decay
pions off the $^3$He nucleus might explain the observed threshold dip
in the production amplitude. Their Monte Carlo simulations suggested
that this would lead to a difference in the measured cross sections
for the $\omega \to \pi^0\pi^+\pi^-$ and $\omega \to \pi^0\gamma$
modes of around 10\% at 1360~MeV. Unfortunately, the uncertainties in
the number of $\omega$ events from the two different channels
obtained in the present experiment are too large to test this
prediction.\vspace{2mm}

Apart from the two points at $\cos\theta_{\omega}^*=-0.7$ and $-0.55$
in Fig.~\ref{fig:ang1360}, which have especially large systematic as
well as statistical uncertainties, the angular distribution is fairly
flat and is in good agreement with the SPESIV point~\cite{wurzinger}.
Excluding these points from a linear fit, the total cross section is
determined to be $\sigma_{\text{tot}}= 84.6\pm4.0\pm 4.8$~nb. To this
must be added the luminosity uncertainty of 29\%.

%
%
\section{Model calculations}
\label{sec:modelcalc}

\begin{figure}[hbt]
\begin{center}
\includegraphics[width=0.8\columnwidth]{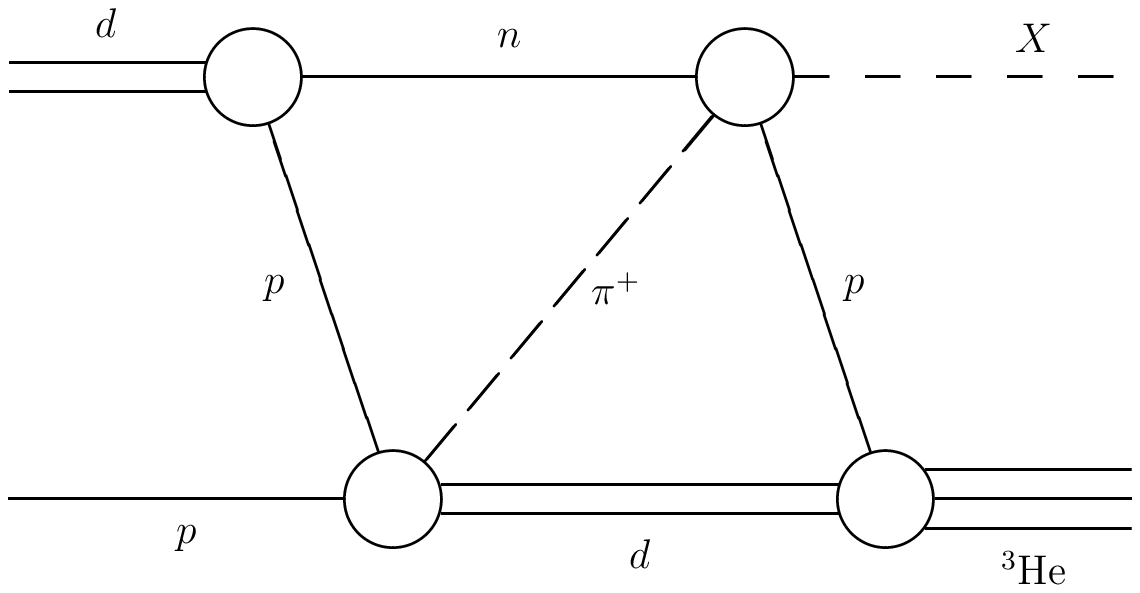}
\caption{Two-step model for the $pd\to{}^3$He$\,X$ reaction where a
virtual pion beam, created through $pp\to d\pi^+$, produces the
observed meson $X$ through a second $\pi^+n\to pX$ reaction. There is
a similar graph with a neutral intermediate pion.}
\label{fig:Feynman}
\end{center}
\end{figure}

The minimum momentum transfer for the $pd\to{}^3$He$\,\omega$
reaction at 1360~MeV is $\sim 1110$~MeV/$c$ and this only drops to
$\sim 935$~MeV/$c$ at 1450~MeV. A direct production mechanism with 
such a large momentum transfer is expected to give vanishingly 
small cross sections. In order to share this large momentum
between the nucleons, Kilian and Nann~\cite{Kilian} suggested a
two-step model, where a virtual pion beam is produced \emph{via} a
$pp\to d\pi^+$ reaction on the target proton, followed by the
production of the observed meson $X$ through the $\pi^+n\to pX$
reaction on the target neutron, as illustrated in
Fig.~\ref{fig:Feynman}. There is an analogous contribution with an
intermediate $\pi^0$.

The quantum-mechanical implementation of Fig.~\ref{fig:Feynman}
describes well the total cross section for the production of the
$\eta$-meson in $pd$ collisions
near-threshold~\cite{gorancolin1,kondrauzi,we3}. The $\eta$ angular
distributions have been measured at several excess energies above
about 20~MeV~\cite{Bilger}. Although it had been claimed their shapes
could be reproduced in a simplified two-step
model~\cite{stenmark2003}, the most realistic implementation of this
approach~\cite{we3_sec} fails to reproduce the data.

Theoretical studies of the $pd\to{}^3$He$\,\omega$ cross section
using a two-step model have been undertaken only close to
threshold~\cite{gorancolin2,kondrauzi} and these underestimate the
experimental data~\cite{wurzinger}. No predictions have yet been made
for the $\omega$ angular distribution.

Following closely the implementation of the two-step approach
described in Ref.~\cite{we3}, we evaluate the $\pi N\to\omega N$
sub-process using the Giessen model~\cite{shklyar}. This is an
effective Lagrangian approach that takes seven coupled channels in
account, \emph{viz.}, $\gamma N$, $\pi N$, $2\pi N$, $\eta N$,
$\omega N$, $K\Lambda$, and $K \Sigma$, for the simultaneous analysis
of all the data up to 2~GeV in terms of 11 isospin-$\frac{1}{2}$
resonances. It was shown in Ref.~\cite{shklyar} that $s$-waves alone
were insufficient to describe the $\pi N\to\omega N$ data even in the
very near-threshold region. The full data set required partial waves
up to $\ell=3$ and their full amplitude analysis has been used as the
basis for the construction of the $\pi N\to\omega N$ $t$-matrix that
is the required input for the $pd\to{}^3$He$\,\omega$ estimation. The
$pp\to d\pi^+$ input was taken from the parameterized $t$-matrix of
the SAID group~\cite{arndt}.

\begin{figure}[hbt]
\begin{center}
\includegraphics[width=0.95\columnwidth]{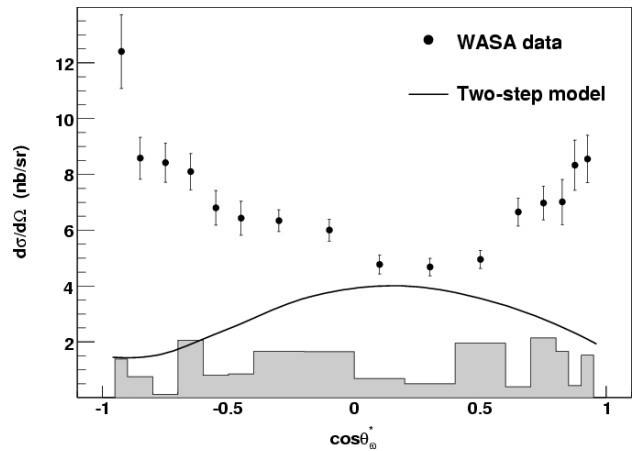}
\caption{Comparison of the WASA data on the $pd\to{}^3$He$\,\omega$
reaction at 1450~MeV (filled circles) with calculations based upon a
two-step approach (solid line). The meaning of the error bars and
histogram is as in Fig.~\ref{fig:ome_ang_exp1450}.}
\label{fig:ome_ang_th1450}
\end{center}
\end{figure}

The results of the calculations performed in the plane wave approach
to the two-step model, are shown in Figs.~\ref{fig:ang1360} and
\ref{fig:ome_ang_th1450} at beam energies of 1360~MeV and 1450~MeV,
respectively, along with the available data. Although the 1450~MeV
predictions are reasonably close to the data in the central region,
the shape is quite different to that of the experiment, giving
forward and backward dips rather than peaks. There are therefore
disagreements of up to an order of magnitude at the extreme angles.
On the other hand, at 1360~MeV the two-step model predicts a fairly
flat distribution with only a slight forward enhancement. However, it
underestimates the cross section by a factor of about five in the
forward region and by even more in the backward direction, which is a
similar discrepancy to that reported in Ref.~\cite{kondrauzi}.

It should be stressed that the $pd\to{}^3$He$\,\omega$ angular
distribution is predicted to be anisotropic at 1360~MeV even if only
$s$-wave are retained for the $\pi N \to \omega N$ vertex. This
anisotropy persists in calculations carried out even closer to the
threshold. It is perhaps interesting to note in this respect that
experimental data on the $pd\to{}^3$He$\,\eta$ reaction very near
threshold show significant anisotropy~\cite{cosy-anke}.

\begin{figure}
\begin{center}
\includegraphics[width=0.95\columnwidth]{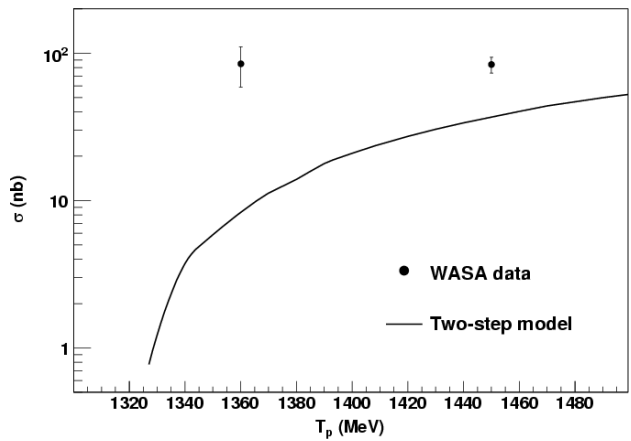}
\caption{The total cross section of $pd\to{}^3$He$\,\omega$ reaction
as a function of the beam kinetic energy. The filled circles are the
two WASA points, where the error bars include the normalization
uncertainties. The solid line shows the estimation within the
two-step model presented in this work.} \label{fig:tot_crossec}
\end{center}
\end{figure}

We compare in Fig.~\ref{fig:tot_crossec} the total cross section
calculated using the two-step model with the WASA data. Since this
theoretical approach underestimates the differential cross sections,
it also underpredicts the total cross sections.

Why does the two-step model fail for $\omega$ production? If the
intermediate pion is taken on-shell, as in the Kilian and Nann
classical model~\cite{Kilian}, the relative velocities of the
deuteron from the $pp\to d\pi^+$ and the proton from the $\pi^+n\to
p\,\omega$ subprocesses are very large at the extreme angles at
1450~MeV. In contrast, at 1360~MeV, the change in the velocity
matching over the full angular region is much less. The cross
sections calculated using the two-step model at 1360~MeV thus
describe the behavior of the experimental angular distribution but
not its magnitude.

The poor velocity matching implies that the intermediate pion is a
long way off its mass shell for both beam energies, unlike the case
of the $pd\to{}^3$He$\,\eta$ reaction at threshold. The predictions
might be improved by using off-shell $t$-matrices for both $pp\to
d\pi^+$ and $\pi^+n\to p\,\omega$. However, it is clear that to
reproduce the data at the extreme angles at 1450~MeV requires
contributions from other diagram(s) since poor velocity matching is a
kinematical effect that depends only upon the assumption of an
intermediate pion. A similar conclusion might be drawn from the
failure of the same model to describe well the $pd\to{}^3$He$\,\eta$
angular distributions away from threshold~\cite{we3_sec}.

%
%
\section{Conclusions}

The differential cross section for the $pd\to{}^3$He$\,\omega$
reaction has been measured over the full angular range at 1450~MeV.
The data show clear anisotropy, with strong rises in both the
backward and forward directions. For the backward-going $\omega$ the
agreement with SPESIII data~\cite{kirchner} is good, but the sharp
forward peaking claimed at SPESIII is not confirmed by our data. The
two-step model fails to describe the angular distribution, giving a
convex rather than the concave shape found in our data

The angular distribution at 1360~MeV is consistent with isotropy and,
within the experimental uncertainties, the differential cross section
for the backward-going $\omega$ agrees with the SPESIV point at
$\cos\theta_{\omega}^* = -1$~\cite{wurzinger}. If the assumptions in
Refs.~\cite{hanhart1,hanhart2} that the SPESIV data are incorrectly
interpreted are right, their point shown in Fig.~\ref{fig:ang1360}
would be higher and the agreement with our data would become worse.

The large angular coverage of the WASA detector allowed values of the
total cross sections at both energies to be extracted for the first
time. At 1450~MeV we find $\sigma_{\text{tot}}=83.6\pm1.5\pm2.2$~nb,
with an additional uncertainty from the normalization of 12\%. The
corresponding number at 1360~MeV is
$\sigma_{\text{tot}}=84.6\pm4.0\pm7.3$~nb, where the additional
normalization uncertainty is 29\%.

The two-step model underestimates the total cross section data by a
factor of about two at 1450~MeV and by a factor of ten at 1360~MeV.
This is probably due to the velocity matching between the proton and
deuteron produced in the two subprocesses being poor for some
$\omega$ angles. Further theoretical work is therefore needed to
describe both $\omega$ and $\eta$ production in $pd$ collisions.\\

%
%
\section*{Acknowledgements}
We wish to thank the personnel at the The Svedberg Laboratory for
their support during the course of the experiment. We are grateful to
Y.~le Bornec for providing us with information regarding
Ref.~\cite{kirchner} and also V. ~Shklyar for supplying the
$t$-matrix element of the Giessen model described in
Ref.~\cite{shklyar}. This work was supported by the European
Community under the ``Structuring the Research Area'' Specific
Program Research Infrastructures Action (Hadron Physics, contract
number RII-cT-204-506078) by the Swedish Research Council, and BMBF
(06TU201). K.P.K.\ wishes to thank the Funda\c{c}\~{a}o para a
Ci\^{e}ncia e a Tecnologia of the Minist\'{e}rio da Ci\^{e}ncia,
Tecnologia e Ensino Superior of Portugal for financial support under
Contract No.\ SFRH/BPD/40309/2007.
%
%


\begin{thebibliography}{35}
%
\bibitem{Binnie}D.~M.~Binnie \emph{et~al.}, Phys.\ Rev.\ D
\textbf{8}, 278 (1973).
%
\bibitem{Keyne} J.~Keyne \emph{et al.}, Phys.\ Rev.\ D \textbf{12},
28 (1976).
%
\bibitem{Karami} H.~Karami \emph{et~al.}, Nucl.\ Phys.\ \textbf{B154}, 503 (1979).
%
\bibitem{plouin} F.~Plouin in
\textit{Production and Decay of Light Mesons}, ed.\ P. Fleury (World
Scientific, Singapore, 1988) p.~114.
%
\bibitem{kirchner} T.~Kirchner, Ph.D. thesis, IPN Orsay, France
(1993).
%
\bibitem{wurzinger} R.~Wurzinger \emph{et~al.},
Phys.\ Rev.\ C \textbf{51}, R443 (1995).
%
\bibitem{hanhart1}
C.~Hanhart and A.~Kudryavtsev, Eur.\ Phys.\ J.\ A \textbf{6}, 325
(1999).
%
\bibitem{hanhart2}
C.~Hanhart, A.~Sibirtsev, and J.~Speth, arXiv:hep-ph/0107245 (2001).
%
\bibitem{gorancolin1} G.~F\"aldt and C.~Wilkin, Nucl.\ Phys.\ \textbf{A587}, 769
(1995).
%
\bibitem{gorancolin2} G.~F\"aldt and C.~Wilkin, Phys.\ Lett.\ B \textbf{354}, 20 (1995).
%
\bibitem{we3}
K.~P.~Khemchandani, N.~G.~Kelkar, and B.~K.~Jain, Nucl.\ Phys.\
\textbf{A708}, 312 (2002).
%
\bibitem{we3_sec}
K.~P.~Khemchandani, N.~G.~Kelkar, and B.~K.~Jain, Phys.\ Rev.\ C
\textbf{68}, 064610 (2003).
%
\bibitem{we4}
N.~J.~Upadhyay, K.~P.~Khemchandani, B.~K.~Jain, and N.~G.~Kelkar
Phys.\ Rev.\ C \textbf{75}, 054002 (2007).
%
\bibitem{kondrauzi}
L.~A.~Kondratyuk and Yu.~N~Uzikov, Acta Physica Polonica B
\textbf{27}, 2977 (1996).
%
\bibitem{stenmark2003}
M.~Stenmark, Phys.\ Rev.\ C \textbf{67}, 034906 (2003).
%
\bibitem{ulla}
U.~Tengblad, G.~F\"aldt, and C.~Wilkin, Eur.\ Phys.\ J.\ A
\textbf{25}, 267 (2005).
%
\bibitem{Karin_PLB} K.~Sch\"{o}nning \emph{et al.}, Phys.\ Lett.\ B \textbf{668}, 258
(2008).
%
\bibitem{MOMO} F.Bellemann \emph{et~al.}, Phys.\ Rev.\ C \textbf{75},
015204 (2007).
%
\bibitem{Zabierowski}
J.~Zabierowski \emph{et~al.}, Physica Scripta \textbf{T99}, 159
(2002).
%
\bibitem{Ekstrom} C.~Ekstr\"om \emph{et~al.}, Physica Scripta \textbf{T99},
169 (2002).
%
\bibitem{Nordhage} {\"O.}~Nordhage, Ph.D. thesis, Uppsala University, Sweden
(2006).
%
\bibitem{birks}
J.~B.~Birks, Proc.\ Phys.\ Soc.\ A \textbf{64}, 874 (1951).
%
\bibitem{jozefPID} J.~Z{\l}oma\'nczuk, \emph{On Particle Identification in the WASA@COSY
Forward Detector}, (2005)
\url{www.tsl.uu.se/~jozef/docs/files/ParticleIdentification.pdf}.
%
\bibitem{nordhage2}
{\"O.}~Nordhage \emph{et~al.}, Nucl.\ Instrum.\ Meth.\ Phys.\ Res.,
Sect.\ A \textbf{569}, 701 (2006).
%
\bibitem{stockholm}
Chr.~Bargholtz \emph{et~al.}, Nucl.\ Instrum.\ Meth.\ Phys.\ Res.,
Sect.\ A \textbf{594}, 339 (2008).
%
\bibitem{berthet}
P.~Berthet \emph{et~al.}, Nucl.\ Phys.\ \textbf{A443}, 589 (1985).
%
\bibitem{karinthesis}
K.~Sch\"onning, Ph.D. thesis, Uppsala University (in preparation).
%
\bibitem{Kilian} K.~Kilian and H.~Nann, AIP Conf.\ Proc.\
\textbf{221}, 185 (1990).
%
\bibitem{Bilger} R.~Bilger \emph{et~al.}, Phys.\ Rev.\ C \textbf{65},
044608 (2002).
%
%
\bibitem{shklyar}
V.~Shklyar, H.~Lenske, U.~Mosel, and G.~Penner, Phys.\ Rev.\ C
\textbf{71}, 055206 (2005).
%
\bibitem{arndt}
R.~A.~Arndt, I.~I.~Strakovsky, R.~L.~Workman, and D.~V.~Bugg,  Phys.\ Rev.\ C \textbf{48}, 1926 (1993).
%
\bibitem{cosy-anke}
T.~Mersmann \emph{et~al.}, Phys.\ Rev.\ Lett.\ \textbf{98}, 242301
(2007); J. Smyrski \emph{et~al.}, Phys.\ Lett.\ B \textbf{649}, 258
(2007).
%
\end{thebibliography}
\end{document}